\documentclass[pra,onecolumn,superscriptaddress,notitlepage]{revtex4-1}
\usepackage{graphicx,amsmath,amsfonts,amssymb,upgreek,txfonts,color}
\usepackage[colorlinks,linkcolor=blue,citecolor=blue,urlcolor=blue,breaklinks=true]{hyperref}

\begin{document}

\title{Optical spatial differentiation with suspended subwavelength gratings}

\author{Alexios Parthenopoulos$^1$, Ali Akbar Darki$^1$, Bjarke R. Jeppesen$^2$, and Aur\'{e}lien Dantan$^1$}

\address{$^1$ Department of Physics and Astronomy, Aarhus University, DK-8000 Aarhus C, Denmark\\
$^2$ Interdisciplinary Nanoscience Center (iNANO), Aarhus University, DK-8000, Aarhus C, Denmark}

\email[Corresponding author: ]{dantan@phys.au.dk} %% email address is required



\begin{abstract}
We demonstrate first- and second-order spatial differentiation of an optical beam transverse profile using thin suspended subwavelength gratings. Highly reflective one-dimensional gratings are patterned on suspended 200 nm-thick silicon nitride membranes using Electron Beam Lithography and plasma etching. The optical transmission of these gratings, designed for illumination with either TM or TE polarized light, are experimentally measured under normal and oblique incidence and found to be in excellent agreement with the predictions of an analytical coupled-mode model as well as Rigorous Coupled Wave Analysis numerical simulations. High quality first- and second-order spatial differentiation of a Gaussian beam are observed in transmission at oblique and normal incidence, respectively. Such easy-to-fabricate, ultrathin and loss-free optical components may be attractive for beam shaping and optical information processing and computing.
\end{abstract}

\maketitle
%%%%%%%%%%%%%%%%%%%%%%%%%%  body  %%%%%%%%%%%%%%%%%%%%%%%%%%

\section{Introduction}

The use of metamaterials and subwavelength-structured devices enables the realization of compact and efficient optical components with tailored optical properties. Of particular interest for all-optical information processing are devices performing spatial or temporal transformations of optical signals. In the spatial domain such ultracompact devices can substantially reduce the footprint of bulky components, such as lenses and filters, and enable integration into complex optical systems. 

Spatial differentiation and integration have been proposed with metasurface-based components~\cite{Silva2014,Pors2015}, as well as with resonant diffractive  devices~\cite{Doskolovich2014,Golovastikov2014,Bykov2014,Ruan2015,Youssefi2016,Hwang2016,Zhu2017,Zangeneh2017,Zangeneh2018,Hwang2018,Guo2018}.
Such diffractive resonant structures can be multilayered systems~\cite{Doskolovich2014,Bykov2014,Ruan2015,Youssefi2016,Zhu2017,Zangeneh2017,Zangeneh2018}, plasmonic structures~\cite{Hwang2016,Hwang2018}, subwavelength gratings~\cite{Golovastikov2014} or photonic crystals~\cite{Guo2018}.

Following the proposal of~\cite{Doskolovich2014}, spatial differentiation was recently implemented with subwavelength TiO$_2$ on quartz gratings~\cite{Bykov2018} and Si on quartz high-contrast subwavelength gratings~\cite{Dong2018}. There, first-order spatial differentiation of the transverse profile of a beam was obtained in transmission by illuminating the grating at oblique incidence at a specific wavelength, determined by the interference between the incident light and a guided-resonant mode in the grating. As pointed out in~\cite{Bykov2018,Dong2018}, such subwavelength structures may be comparatively easier to design and fabricate than metasurfaces and do not require additional lens or filter to perform the required spatial transformations.

Building upon these advances, we demonstrate here both first- and second-order spatial differentiation of an optical beam transverse profile using thin (200 nm) silicon nitride suspended subwavelength gratings. We make use of the recently demonstrated recipe for directly patterning commercial high quality suspended Si$_3$N$_4$ films with one-dimensional subwavelength gratings~\cite{Nair2019}. In contrast with~\cite{Nair2019}, much larger gratings---(200 $\mu$m)$^2$ {\it vs} (50 $\mu$m)$^2$---are realized, allowing for operating with larger beams, thus minimizing focusing and finite size effects and thereby acheiving high reflectivity (>95\%)~\cite{Parthenopoulos2020}. Gratings with different periods/fill factors are fabricated to observe TE or TM resonances in 920-960 nm range. Their transmission spectra under monochromatic illumination at different incidence angles are measured and resonances with different parity guided modes are observed. Very good agreement between the predictions of full Rigorous Coupled Wave Analysis simulations and the simple coupled-mode model of Bykov {\it et al.}~\cite{Bykov2015,Bykov2017} is obtained. High quality first- and second-order spatial differentiation of a Gaussian beam transverse profile is observed at oblique and normal incidence, respectively, for either TM or TE incident polarized light.

Such easy-to-fabricate, ultrathin and loss-free components are attractive for beam shaping and optical information processing and computing and add to the panoply of applications of subwavelength grating-based devices~\cite{Chang-Hasnain2012,Quaranta2018}. Moreover, they may enable new applications of suspended Si$_3$N$_4$ thin films, possessing both high optical and mechanical quality and which are widely used within optomechanics~\cite{Thompson2008,Kemiktarak2012,Bui2012,Kemiktarak2012NJP,Xuereb2012,Norte2016,Reinhardt2016,Li2016,Chen2017,Nair2017,Moura2018,Naesby2018,Gartner2018,Cernotik2019,Dantan2020,Manjeshwar2020}.

%%%%%%%%%%%%%%%%%%%%%%%%%%%%%%%%%%%

\section{Theoretical model and simulations}

\subsection{Coupled mode model}

\begin{figure}
\centering
\includegraphics[width=0.5\textwidth]{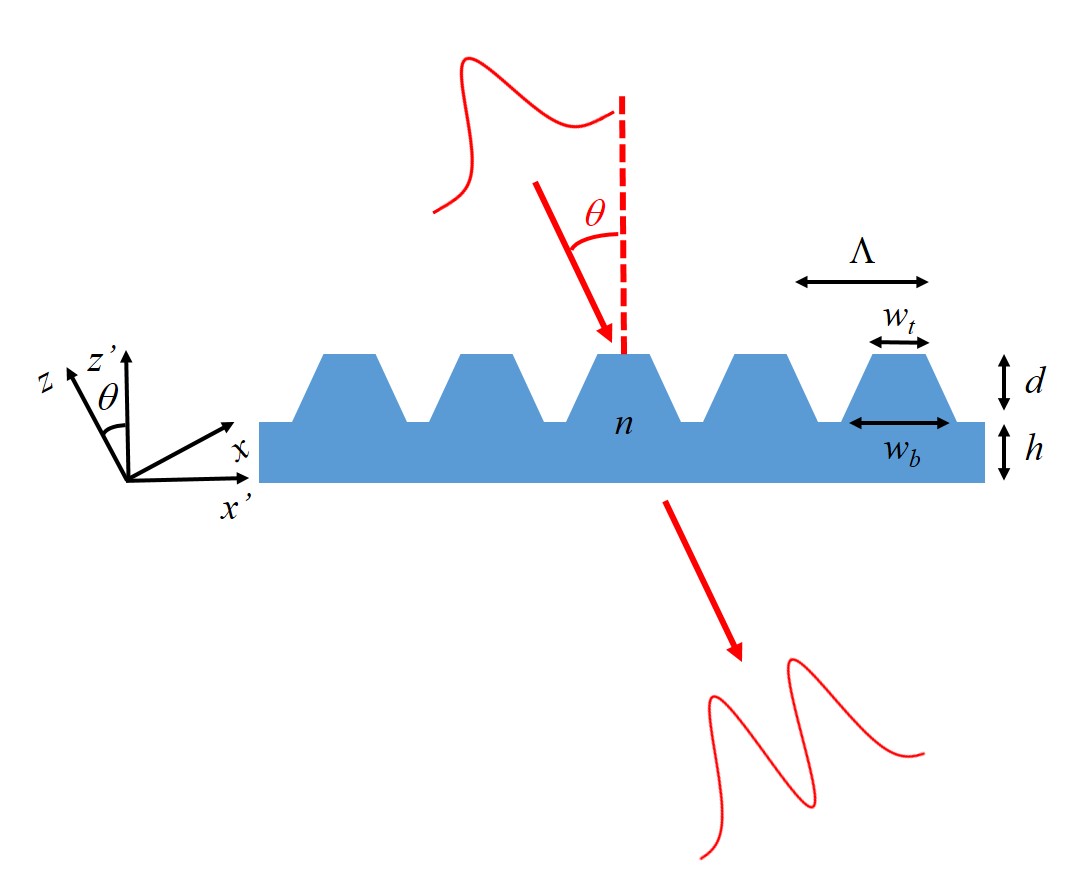}
\caption{Spatial differentiation with a suspended subwavelength grating: an incident beam linearly polarized in the $(xy)$-plane impinges on a one-dimensional subwavelength grating periodic in the $x'$-direction, the incident beam propagation direction $z$ making with an angle $\theta$ with the direction $z'$ normal to the grating plane. For specific incidence angles/wavelengths of the incident light beam the profile of the transmitted zeroth-order beam is that of the incident beam, spatially differentiated with respect to the $x$-coordinate.}
\label{fig:grating_schematic}
\end{figure}

We follow Bykov {\it et al.}'s approach and notations~\cite{Golovastikov2014,Bykov2018} and consider an infinite one-dimensional grating with period $\Lambda$ illuminated by a monochromatic light beam with wavelength $\lambda$ and transverse profile amplitude $P_{\textrm{inc}}(x,y)$ (Fig.~\ref{fig:grating_schematic}). The incident beam is linearly polarized in the $(xy)$-plane and its propagation direction makes an angle $\theta$ with the direction normal to the grating plane. Due to the subwavelength nature of the grating, the transmitted beam is in the zeroth diffraction order and its profile, denoted by $P_{\textrm{tr}}(x,y)$, is obtained from that of the incident beam using a transfer function~\cite{Golovastikov2014,Bykov2018}
\begin{equation}
H(k_x,k_y)\simeq T\left(k_x\cos\theta+\sqrt{k_0^2-k_x^2-k_y^2}\sin\theta,k_y\right),
\label{eq:H1}
\end{equation}
where $k_0=2\pi/\lambda$ is the free-space wave number and $T(k_x',k_y)$ is the transmission coefficient of the grating, where $k_{x'}$ is the component of the wave vector in the grating coordinate system (Fig.~\ref{fig:grating_schematic}). An elegant coupled mode model was put forward by Bykov {\it et al.} in~\cite{Bykov2017} to derive a generic approximate expression for this transmission coefficient, which, in the vicinity of the guided mode resonance and for small angles of incidence, reads
\begin{equation}
T(k_{x'},k_y)=t\frac{v_g^2k_{x'}^2-(\omega-\omega_0-\eta_0k_y^2)(\omega-\omega_2-\eta_2k_y^2)}{v_g^2k_{x'}^2-(\omega-\tilde{\omega}_1-\eta_1k_y^2)(\omega-\omega_2-\eta_2k_y^2)},
\label{eq:Tgen}
\end{equation}
where $\omega=2\pi c/\lambda$ is the light field frequency, $c$ being the speed of light, and $t$, $v_g$, $\omega_0$, $\tilde{\omega}_1$, $\omega_{2}$ and $\eta_i$ ($i=0,1,2$) are parameters depending on the grating geometry and refractive index.

Choosing $k_{y0}=0$, $k_{x'0}=v_g^{-1}[(\omega-\omega_0)(\omega-\omega_2)]^{1/2}$ results in a vanishing transmission coefficient, $T(k_{x'0},k_{y0})=0$. Assuming a weakly focused beam with similar spectral widths in the $x$ and $y$ directions and expanding the transmission coefficient in the vicinity of this point at lowest order in $k_{x'}$, $k_y$ yields
\begin{equation}
T(k_{x'},k_y)\simeq\alpha(k_x-k_{x'0}),
\label{eq:Tlinear}
\end{equation}
where $\alpha$ can be obtained from the phenomenological parameters introduced in Eq.~(\ref{eq:Tgen}). Since, for a weakly focused beam, Eq.~(\ref{eq:H1}) becomes \begin{equation}H(k_x,k_y)\simeq T(k_x\cos\theta+k_0\sin\theta,k_y),\end{equation} choosing the angle of incidence $\theta$ such that $\sin\theta=k_{x'0}/k_0$ allows for reexpressing the transfer function of the grating as that of a spatial differentiator
\begin{equation}
H(k_x,k_y)\simeq ik_x,
\end{equation}
up to a proportionality constant $-i\alpha\cos\theta$.

At normal incidence ($\theta=0$), the transfer function is not linear, but quadratic in $k_{x'}$. This can be exploited to perform a second order differentiation of the beam profile, as suggested in~\cite{Doskolovich2014} and as demonstrated in Sec.~\ref{sec:diff}.

\subsection{RCWA simulations}

\begin{figure}
\centering
\includegraphics[width=0.5\textwidth]{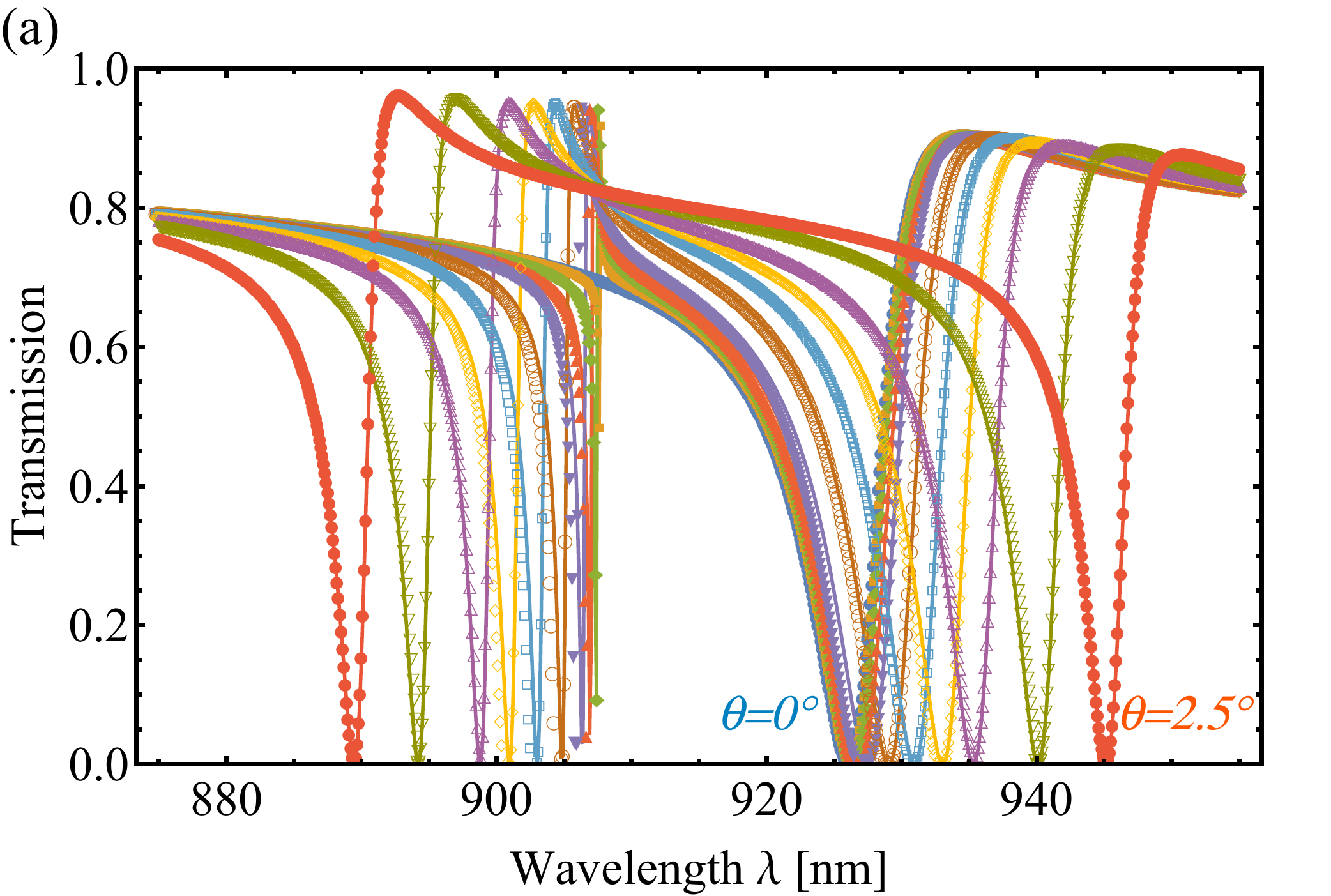}\includegraphics[width=0.5\textwidth]{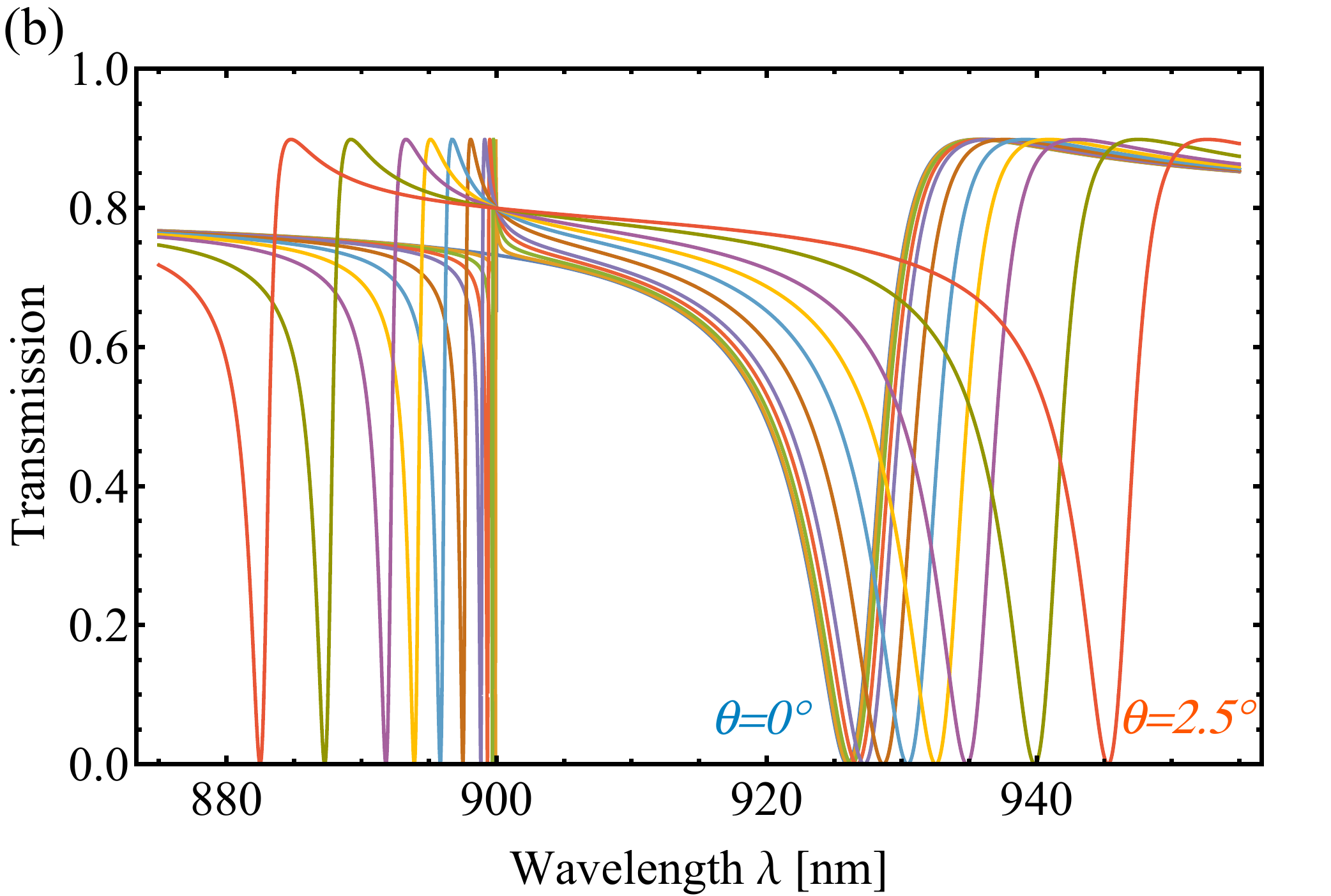}
\caption{(a) Simulated RCWA transmission spectra of a grating showing guide mode resonances under TM polarized light illumination (see text for parameters) for different angles of incidence ($\theta=0^\circ,0.125^\circ,0.25^\circ,0.375^\circ,0.5^\circ,0.75^\circ,1^\circ,1.25^\circ,1.5^\circ,2^\circ,2.5^\circ$). (b) Results of fits of the simulated spectra to the coupled mode model [Eq.~(\ref{eq:Tlambdatheta})].}
\label{fig:TMsim}
\end{figure}

\begin{figure}
\centering
\includegraphics[width=0.5\textwidth]{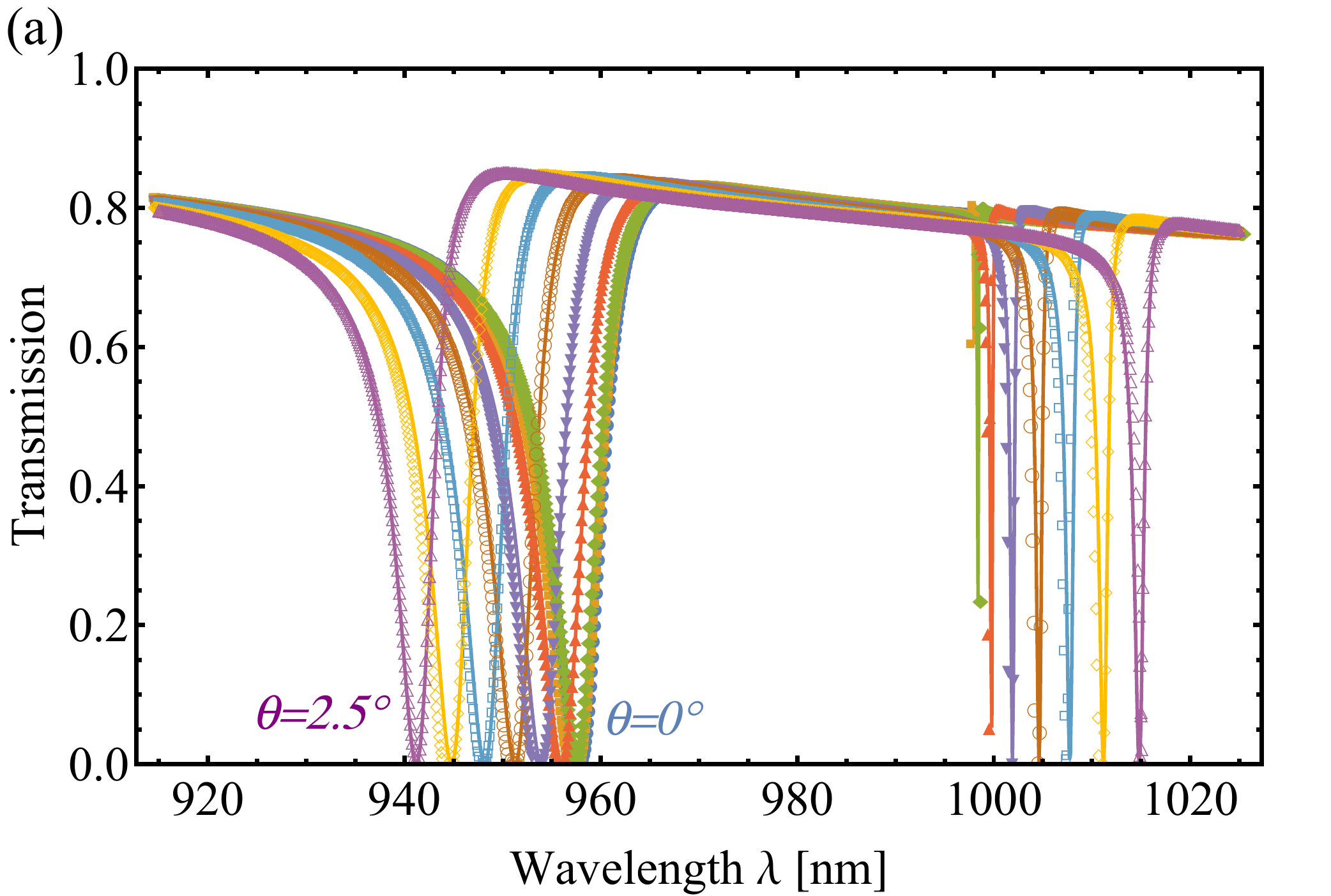}\includegraphics[width=0.5\textwidth]{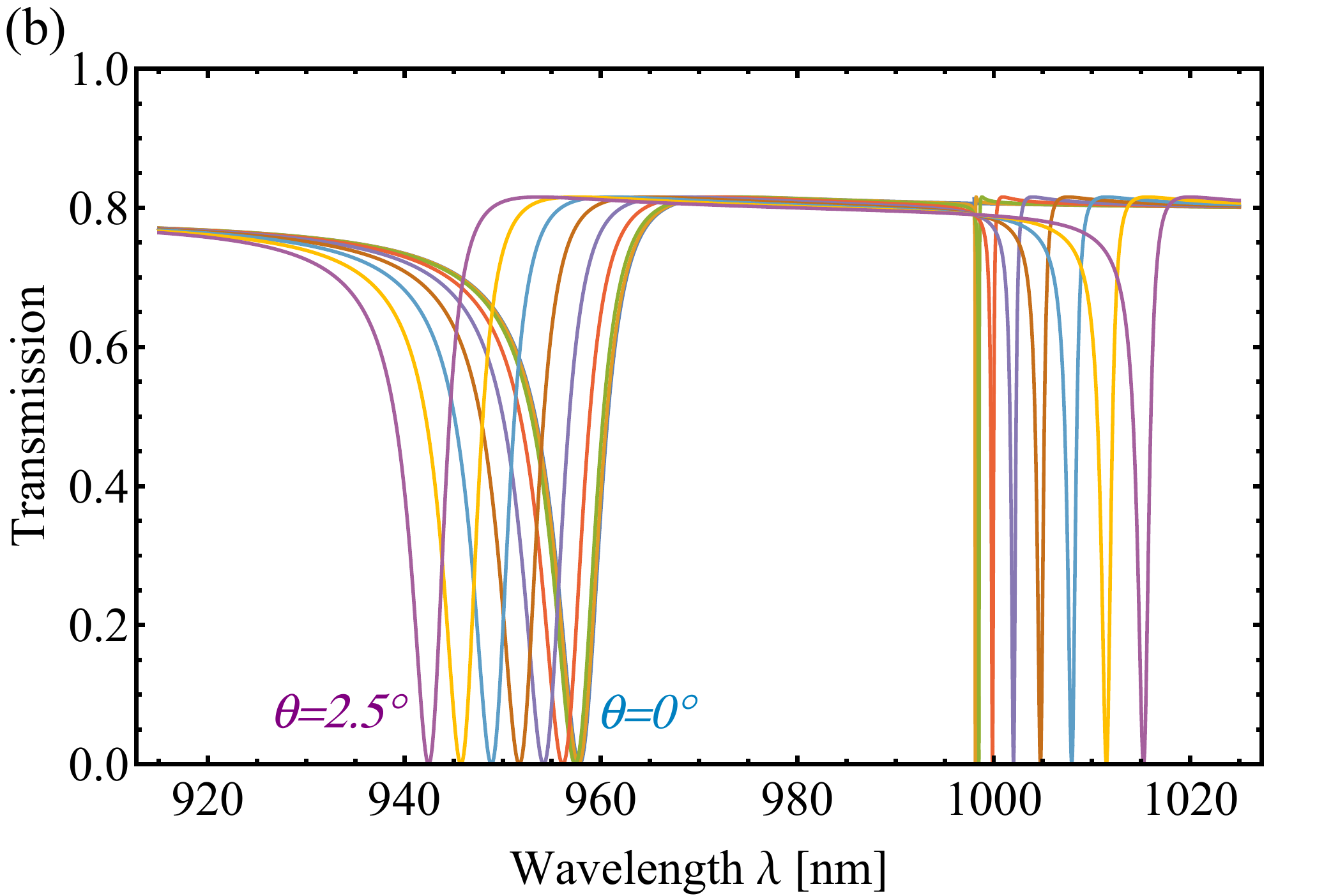}
\caption{(a) Simulated RCWA transmission spectra of a grating showing guide mode resonances under TE polarized light illumination (see text for parameters) for different angles of incidence ($\theta=0^\circ,0.25^\circ,0.5^\circ,1^\circ,1.5^\circ,2^\circ,2.5^\circ,3^\circ,3.5^\circ$). (b) Results of fits of the simulated spectra to the coupled mode model [Eq.~(\ref{eq:Tlambdatheta})].}
\label{fig:TEsim}
\end{figure}

To illustrate the accuracy of this coupled mode model and prepare the ground for the experimental observations of the following sections, we numerically simulated the transmission spectra of infinite one-dimensional gratings with parameters similar to those whose fabrication will be detailed in Sec.~\ref{sec:fab} using full RCWA calculations~\cite{MIST}. We thus considered infinite lossless gratings made of silicon nitride (refractive index 2.0) and whose profile is shown in Fig.~\ref{fig:grating_schematic}. The grating fingers are trapezoidal with top and bottom widths $w_t$ and $w_b$, respectively, and have a height $d$. The underlying silicon nitride layer has a thickness $h$. We assume infinite plane wave illumination with light impinging on the grating, as depicted in Fig.~\ref{fig:grating_schematic}, and with either TM or TE polarization. Figures~\ref{fig:TMsim} and \ref{fig:TEsim} show the simulated (intensity) transmission spectra for different incidence angles of two gratings designed to observe guided mode resonances in the range 915-975 nm and whose fabrication and characterization are detailed in Sec.~\ref{sec:fab}. The simulated spectra are calculated using the MIST software~\cite{MIST}, discretizing the medium in 20 layers and using a 25 mode basis. The parameters of the grating whose spectra are shown in Fig.~\ref{fig:TMsim} are $\Lambda=802.3$ nm, $w_t=458.2$ nm, $w_b=607.2$ nm, $d=93.7$ nm and $h=98.8$ nm and the light polarization is TM, while those of the grating whose spectra are shown in Fig.~\ref{fig:TEsim} are $\Lambda=647$ nm, $w_t=462.2$ nm, $w_b=599.8$ nm, $d=94.9$ nm and $h=96.3$ nm and the light polarization is TE. The essential difference between these two gratings is thus a shorter period for the "TE grating" (and thereby a higher fill factor) as compared to the "TM grating". As discussed in the next section, this choice is merely motivated by the possibility of being able to observe the TM or TE resonances within the available tuning range of the laser used for characterizing the samples.

These spectra can be straightforwardly interpreted in terms of the parameters of the coupled mode model introduced in the previous section. Indeed, rewriting $T(k_{x'0},0)$ in the small incidence approximation yields a transmission coefficient equal to
\begin{equation}
\mathcal{T}(\lambda,\theta)=t\frac{\nu^2\theta^2-(\omega-\omega_0)(\omega-\omega_2)}{\nu^2\theta^2-(\omega-\tilde{\omega}_1)(\omega-\omega_2)}.
\label{eq:Tlambdatheta}
\end{equation}
As shown in~\cite{Bykov2015,Bykov2017}, $\nu$, $\omega_0$ and $\omega_2$ are real parameters for a lossless grating. $\omega_0$ thus represents the light frequency at which the transmission vanishes at normal incidence. $\omega_2$ represents the resonance frequency of the odd guided mode. $\tilde{\omega}_1$ is a complex number, which we choose to write as $\tilde{\omega}_1=\omega_1+i\gamma$, where $\omega_1$ is the resonance frequency of the even guided mode and $\gamma$ its width. $\nu$ is a constant depending on the mode group velocity $v_g$ and determines the shifts of the guided mode resonances with the incidence angle. $|t|^2$ is the intensity transmission coefficient far from the resonances. Note that, at normal incidence ($\theta=0$), the transmission coefficient simply takes the well-known form
\begin{equation}
\mathcal{T}(\omega,0)=t\frac{\omega-\omega_0}{\omega-\omega_1-i\gamma}
\end{equation}
which yields a single Fano resonance in the transmission spectrum, as the incident wavefront symmetry only allows for exciting the even guided mode at normal incidence~\cite{Fan2002,Fan2003,Bykov2015,Bykov2017}. At oblique incidence, the odd guided mode can be excited and a second Fano resonance appears. It is easy to see from Eq.~(\ref{eq:Tlambdatheta}) that both resonances are shifted further apart---quadratically with $\theta$---as the incidence angle is increased.

The results of fits of the simulated spectra of Figs.~\ref{fig:TMsim}(a) and \ref{fig:TEsim}(a) to the coupled model [Eq.~(\ref{eq:Tlambdatheta})] are shown in Figs.~\ref{fig:TMsim}(b) and \ref{fig:TEsim}(b), respectively. For completeness, fitting first the normal incidence spectrum and subsequently performing a global fit to the various incidence angle spectra yields the following parameters (expressed in wavelength units): $\lambda_0=926.0$ nm, $\lambda_1=927.0$ nm, $\lambda_2=900.0$ nm, $\gamma_\lambda=\gamma(2\pi c/\lambda_0^2)=3.0$ nm, $\nu/c=8.6\times 10^{-5}$ (nm.deg)$^{-1}$ and $|t|^2=0.79$ for the "TM grating" of Fig.~\ref{fig:TMsim}, and $\lambda_0=957.8$ nm, $\lambda_1=958.3$ nm, $\lambda_2=998.0$ nm, $\gamma_\lambda=2.7$ nm, $\nu/c=5.7\times 10^{-5}$ (nm.deg)$^{-1}$ and $|t|^2=0.79$ for the "TE grating" of Fig.~\ref{fig:TEsim}. The shapes and widths of the resonances, as well as the magnitude of the resonance shifts with the incidence angle, are roughly similar for both gratings, but, since $\omega_1>\omega_2$ for the TM grating, the even mode resonance is shifted towards higher wavelengths, whereas the opposite occurs for the TE grating. 

A very good overall agreement is observed, showing that the coupled mode model captures well the salient features of the interferences between the incident and the guided modes. Such an agreement was already evidenced in~\cite{Bykov2015} in the case of a deeply subwavelength guided-mode resonant filter;  remarkably, it also holds in the case of the gratings studied here, which operate in an intermediate regime between that of deep, high-contrast gratings and that of shallow, weakly modulated subwavelength gratings.

%%%%%%%%%%%%%%%%%%%%%%%%%%%%%%%%%%%

\section{Fabrication and characterization}
\label{sec:fab}

\begin{figure}[h]
\centering
\includegraphics[width=\textwidth]{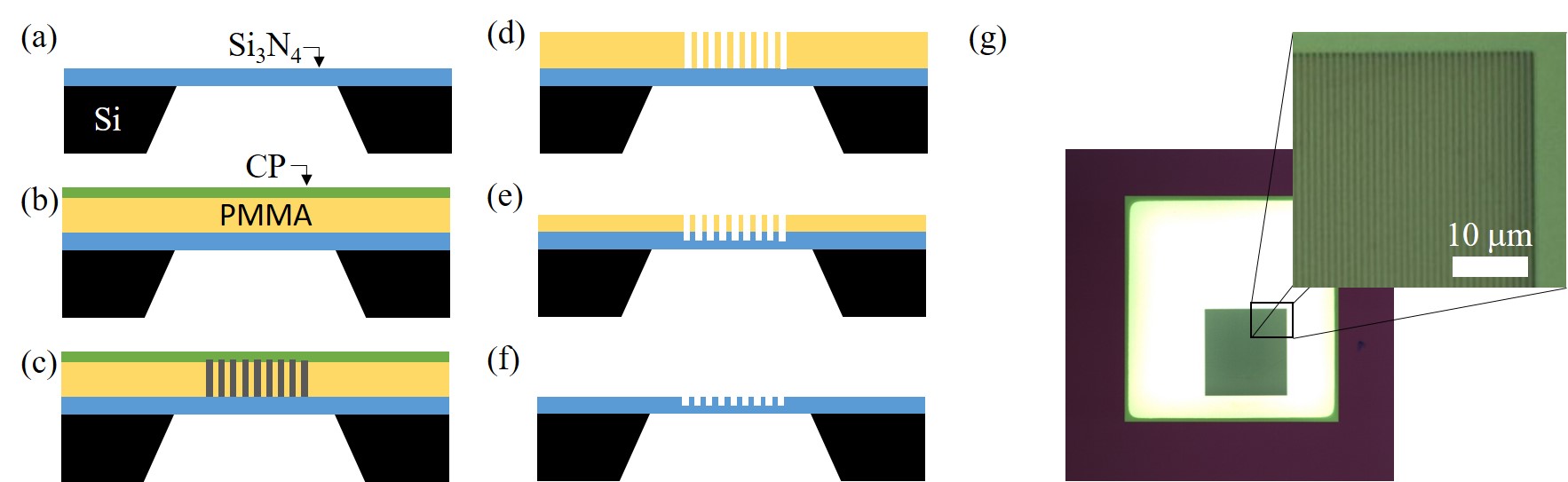}
\caption{Fabrication process: (a) Suspended Si$_3$N$_4$ membrane on Si. (b) Coating with PMMA and conductive polymer layer. (c) Electron Beam Lithography. (d) Development. (e) Dry etching. (f) Suspended patterned membrane. (g) Microscope image of the membrane (white color) patterned with a "TM grating" (green color and zoom-in).}
\label{fig:fabrication}
\end{figure}

To fabricate the subwavelength gratings, a recipe similar to that demonstrated in~\cite{Nair2019} was used. We start with commercial (Norcada Inc., Canada), high tensile stress ($\sim$ GPa), stochiometric silicon nitride suspended thin films. The silicon nitride films are 200 nm-thick and deposited on a 5 mm-square, 200 $\mu$m-thick silicon frame. The lateral dimension of the suspended square membrane is 500 $\mu$m and the lateral size of the patterned grating area is 200 $\mu$m.

The subwavelength grating structures are realized following the steps depicted in Fig.~\ref{fig:fabrication}. After oxygen plasma cleaning the samples are spin-coated with a 550 nm-thick layer of 7\% 950k PMMA and a 50 nm-thick conductive polymer (SX-AR-PC 5000). A square grating mask is written by EBL (dose 310 $\mu$C/cm$^2$, acceleration voltage 30 kV). The conductive polymer layer is subsequently removed by immersion in deionized water and the PMMA resist is developed in a solution of 3:7 H$_2$O:IPA for 1 min and stopped by direct dipping into pure IPA for 30 s. The sample is then etched in a STS Pegasus ICP DRIE system using reactive ion etching with C$_4$F$_8$ (flow rate 59 sccm) and SF$_6$ (flow rate 36 sccm) for 95 s at an rf power 800 W. The PMMA layer is removed in acetone and the sample is cleaned and dried with N$_2$. 

An optical microscope picture of the obtained "TM grating" structure is shown in Fig.~\ref{fig:fabrication}(g). Homogenous grating structures with trapezoidal fingers with a depth corresponding to roughly half the slab thickness are obtained. Instead of the (destructive) Focused Ion Beam cutting method used in~\cite{Nair2019}, the transverse profile of the subwavelength gratings is noninvasively determined here using Atomic Force Microscopy scans, allowing us to accurately extract the geometrical grating parameters (period, finger top and bottom widths, finger depth)~\cite{Darki2020}. The (unpatterned) thin film refractive index and thickness are determined independently by ellipsometry or white light interferometry~\cite{Nair2017}. This characterization provides the input parameters used in the simulations of the previous section.

\begin{figure}[h]
\centering
\includegraphics[width=\textwidth]{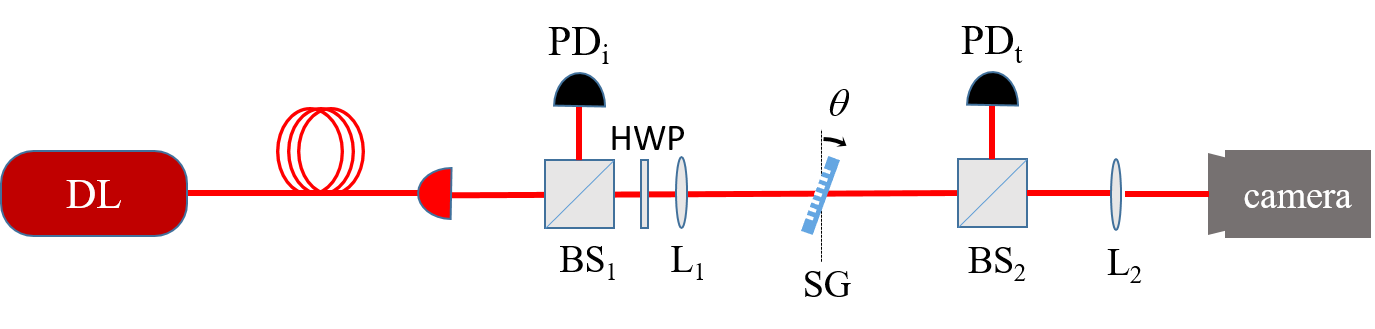}
\caption{Optical characterization setup. DL: diode laser, BS: beamsplitter, PD: photodiode, HWP: halfwave plate, L: lens, SG: subwavelength grating.}
\label{fig:setup}
\end{figure}

In order to characterize the optical transmission properties of the gratings under monochromatic illumination, monochromatic light from a tunable diode laser (Toptica DLC Pro 915-985 nm) is coupled into a single-mode polarization maintaining fiber and approximately half of the light exiting the fiber is weakly focused using an achromatic 75 mm-focal length doublet (L$_1$) onto the membrane which rests on a mount with adjustable tilt. The polarization of the incident light is set by an achromatic halfwave plate (Thorlabs AHWP05M-980). Part of the transmitted light is sent to a photodetector using a beamsplitter (BS$_2$), while the light reflected by the second beamsplitter is collected by a 75 mm-focal length lens (L$_2$) and a CMOS camera (Thorlabs DCC1545) with an Edmund Optics 83-891 objective, in order to provide an image of the transverse profile of the beam at the membrane with a magnification of approximately 3. The transmitted signal measured by the photodetector (PD$_\textrm{t}$), in combination with the monitored incident power (PD$_\textrm{i}$), is used to determine the normalized transmission of the sample, by performing scans of the laser wavelength with and without membrane.

The larger size (and high quality) of the subwavelength grating structures realized---(200 $\mu$m)$^2$ versus (50 $\mu$m)$^2$ in~\cite{Nair2019}---allows for reducing collimation and finite size effects, as compared to~\cite{Nair2019}, by operating with large beam waists ($\sim65-70$ $\mu$m beam waist radius here) and thereby achieving high reflectivity (>95\%)~\cite{Parthenopoulos2020}.

%%%%%%%%%%%%%%%%%%%%%%%%%%%%%%%%%%%

\section{Spatial differentiation}
\label{sec:diff}

%\subsection{"TM grating"}

\begin{figure}[h]
\centering
\includegraphics[width=0.75\textwidth]{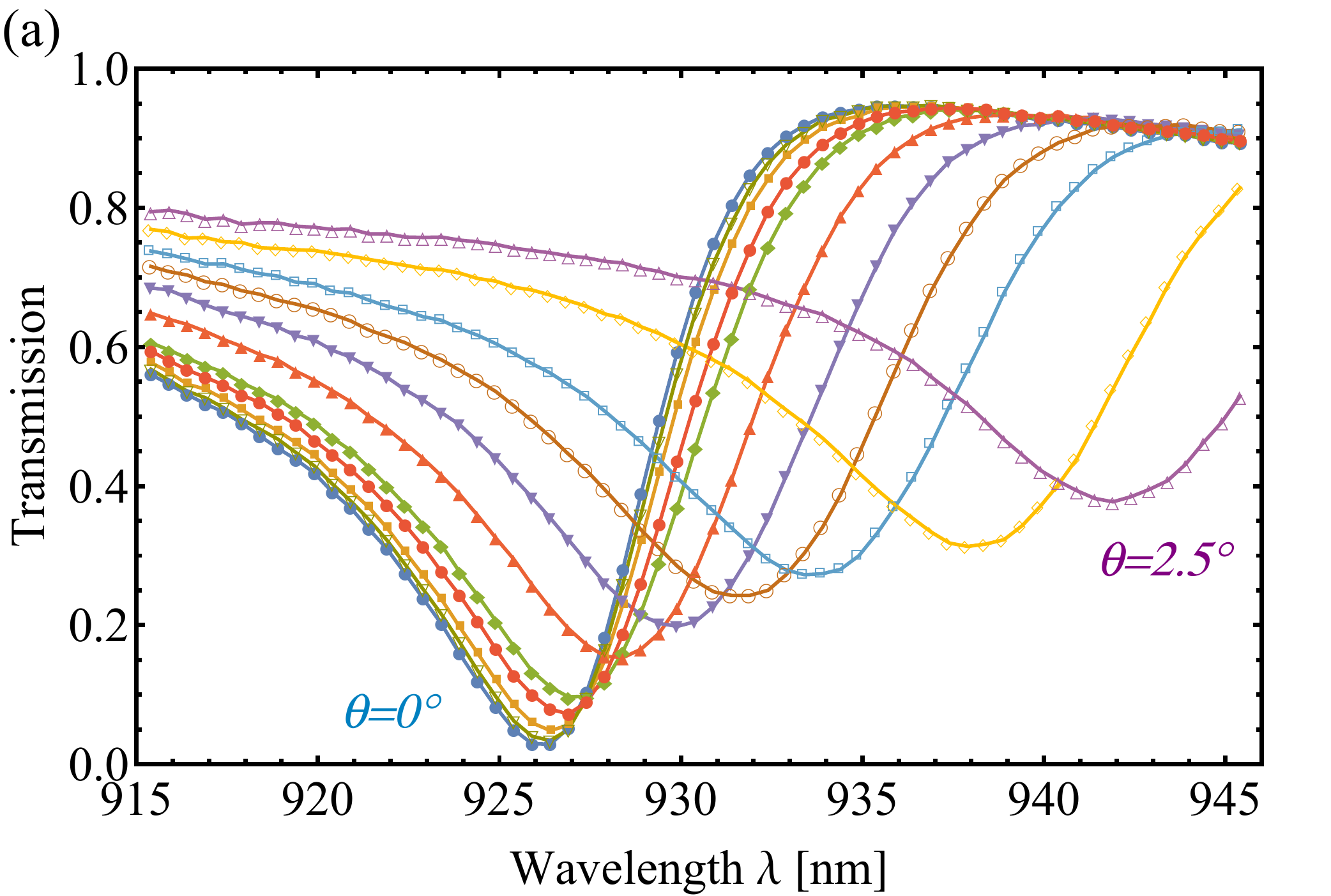}\\
\includegraphics[width=0.75\textwidth]{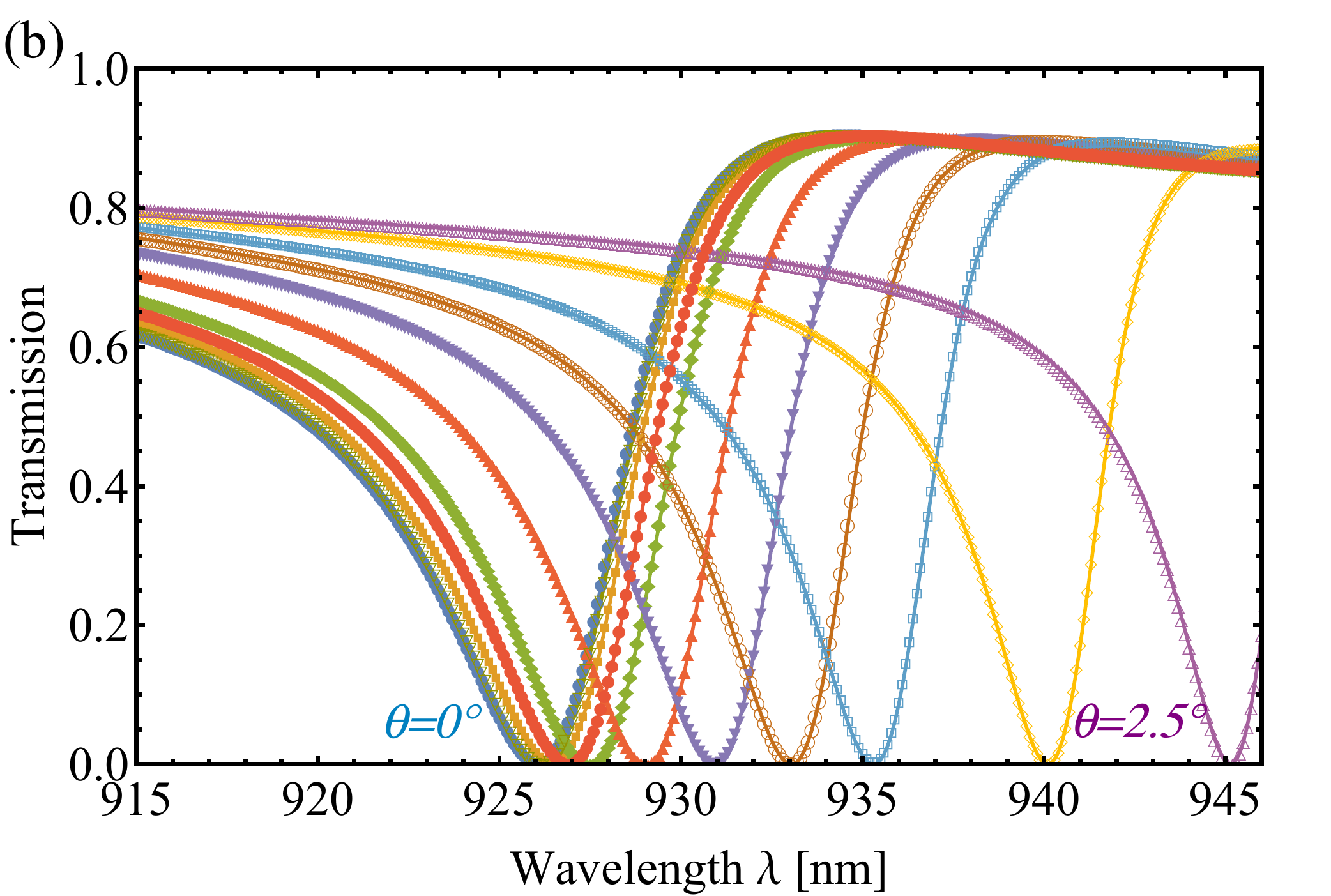}\\
\includegraphics[width=\textwidth]{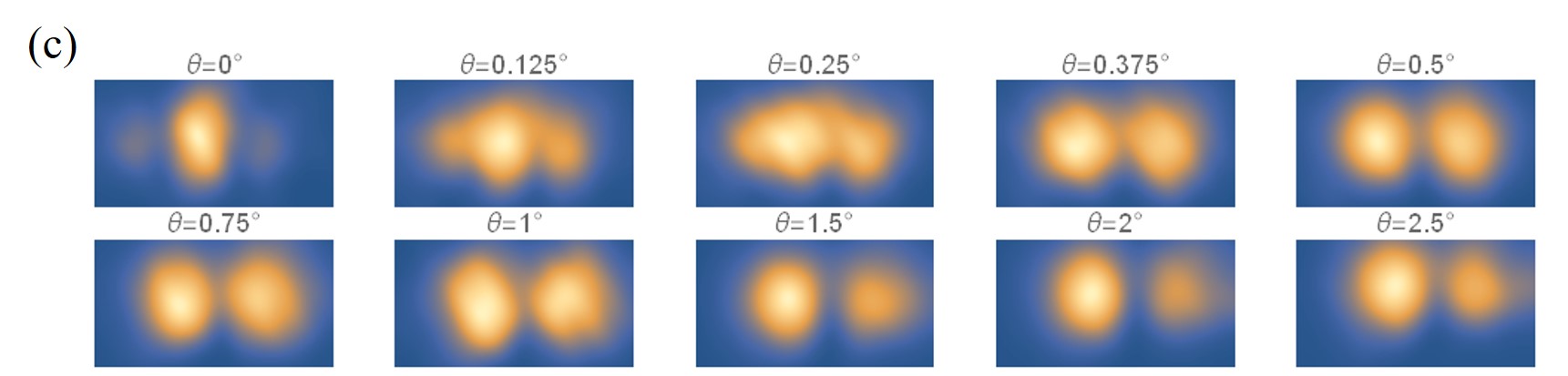}
\caption{(a) Measured transmission spectra for the "TM grating" and illumination at different incidence angles ($\theta=0^\circ,0.125^\circ,0.25^\circ,0.375^\circ,0.5^\circ,0.75^\circ,1^\circ,1.25^\circ,1.5^\circ,2^\circ,2.5^\circ$). (b) RCWA simulated transmission spectra for the "TM grating" under plane-wave illumination at the same incidence angles. (c) Measured images of the transmitted beam transverse profile at the  minimum transmission wavelength for each incidence angle.}
\label{fig:TMspectra}
\end{figure}

\begin{figure}[h]
\centering
\includegraphics[width=0.75\textwidth]{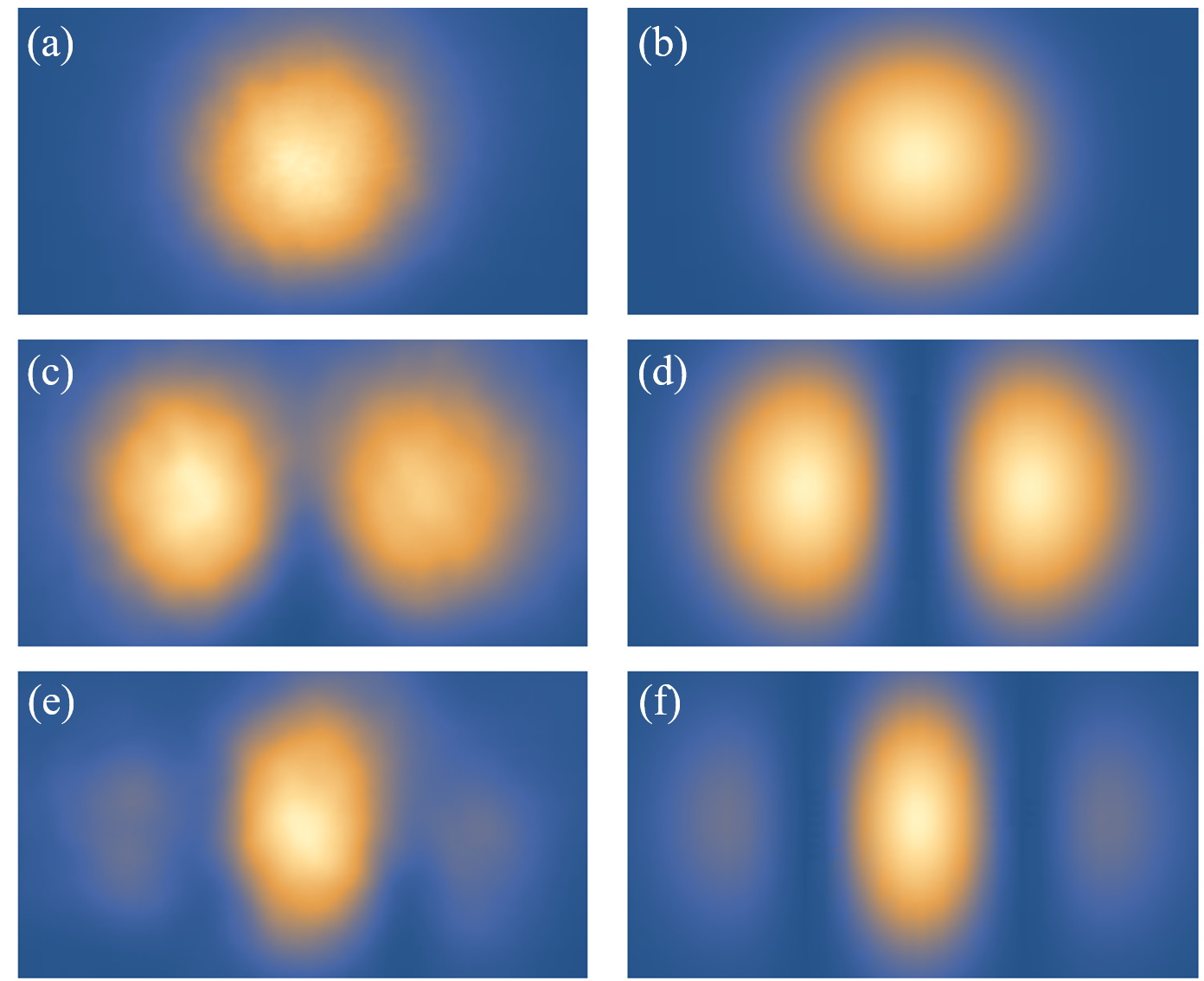}\\
\includegraphics[width=0.75\textwidth]{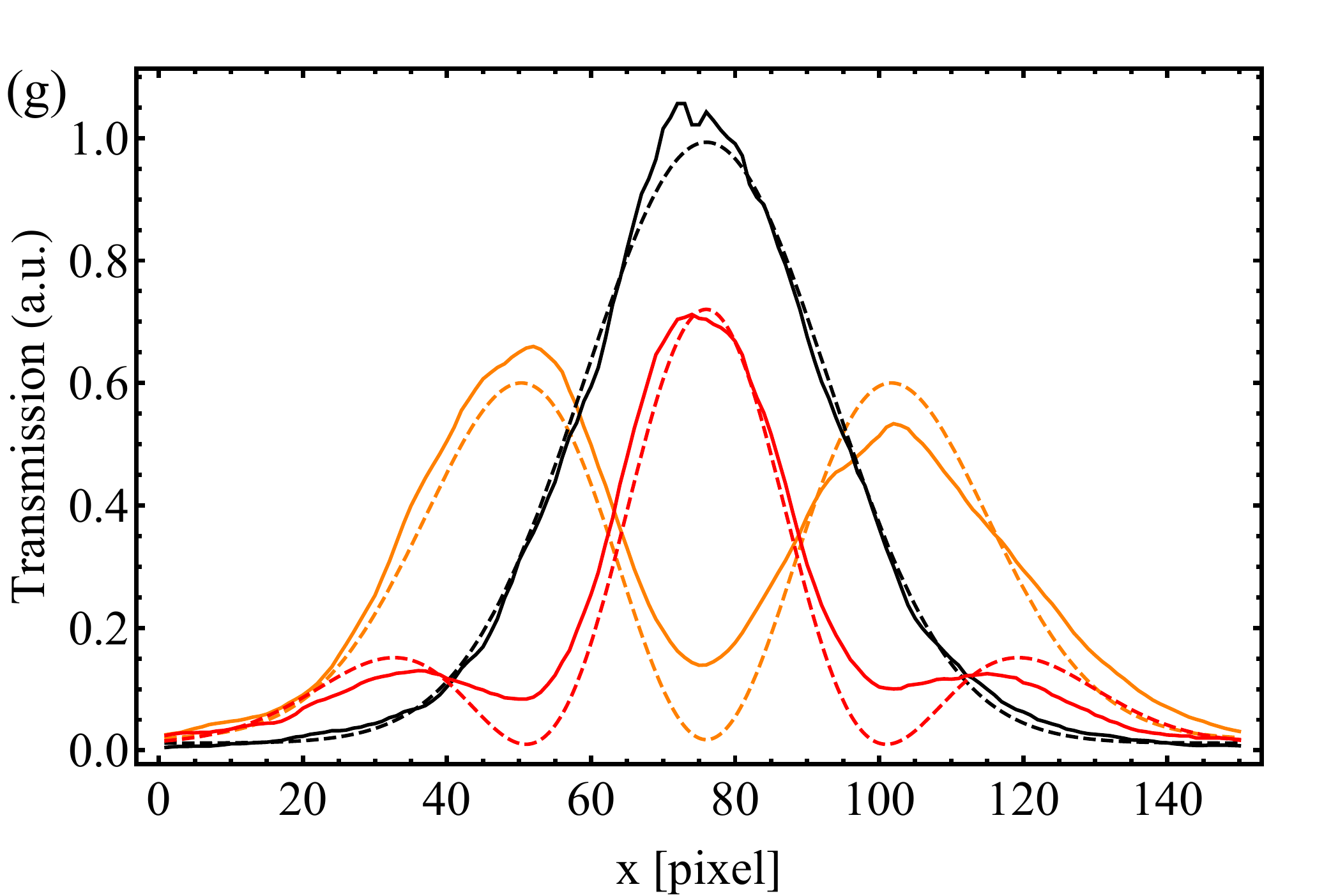}
\caption{Images of the transverse profile of the incident beam (a) and the transmitted beam by the "TM grating" at the minimum transmission wavelength for $1^\circ$ (c) and $0^\circ$ (e) angles of incidence. (b), (d) and (f) show the results of fits to a non-, first-order and second-order differentiated Gaussian beam profiles. (g) Transverse profile of the central part of the different beams in the $x$ direction (plain), together with the corresponding fit results (dashed).}
\label{fig:TMmodes}
\end{figure}

We now investigate the spatial differentiation of a Gaussian beam using the "TM grating" fabricated in the previous section. The beam exiting the single mode fiber is to a very good approximation in a fundamental Hermite-Gauss mode. Figure~\ref{fig:TMmodes}(a) shows the transverse profile of the incident beam imaged at the membrane position, while Fig.~\ref{fig:TMmodes}(c) shows the result of a fit to a Gaussian profile $\sim \exp[-2(x^2+y^2)/w_0^2]$, yielding a $\chi^2$ fit coefficient of 0.996. Figures~\ref{fig:TMspectra}(a) and (c) show the transmission spectra through the grating at various incidence angles between $0^\circ$ and $3.5^\circ$, as well images of the transverse profile of the transmitted beams taken at the corresponding minimum transmission wavelength.\\

At normal incidence a single Fano resonance is observed at 925.8 nm, where the minimum transmission is $\sim$3.4\%). At oblique incidence, the Fano resonance is shifted towards higher wavelengths, as expected from the theoretical predictions; the corresponding simulated spectra of Fig.~\ref{fig:TMsim}(a) are shown again in Fig.~\ref{fig:TMspectra}(b) for reference. As the incidence angle increases the width of the resonance and the minimum transmission increase as well, due to collimation effects (which are not taken into account in the RCWA simulations assuming plane wave illumination).\\ 

Second-order spatial differentiation is observed at normal incidence, while first-order spatial differentiation is observed at oblique incidence, with a rapid transition of the order of few tens of a degree. As examples, Figs.~\ref{fig:TMmodes}(c) and (e) show images of the transverse profile at the minimum transmission wavelength for $\theta=1^\circ$ and $\theta=0^\circ$, respectively. Figures~\ref{fig:TMmodes}(d) and (f) show the results of fits to first-order differentiated ($\sim x^2\exp[-2(x^2+y^2)/w_0^2] $) and second-order differentiated ($(1-2x^2/w_0^2)^2 \exp[-2(x^2+y^2)/w_0^2] $) Gaussian beam intensity profiles, respectively. The resulting $\chi^2$ fit coefficients are respectively 0.963 and 0.940, showing the good quality of the spatial differentiation. The transverse profiles of the central part of the different beams in the $x$ direction are shown in Fig.~\ref{fig:TMmodes}(g), together with the corresponding fit results.\\

Similar experiments were performed with the "TE grating" previously described and similar results in excellent agreement with the theoretical predictions were obtained, as shown in the Appendix.

%%%%%%%%%%%%%%%%%%%%%%%%%%%%%%%%%%%

\section{Conclusion}

In-depth investigations of spatial differentiation of optical beams by thin dielectric subwavelength gratings under various illumination (incidence angle, polarization) conditions were carried out. High quality first- and second-order spatial differentiation of a Gaussian beam was demonstrated at oblique and normal incidence, respectively. The experimental observations are in excellent agreement with the predictions of full RCWA simulations and the generic coupled-mode model of Bykov {\it et al.}, which is used as a basis for calculating the transfer functions of the spatial differentiators. The direct patterning of suspended, essentially loss-free, commercial silicon nitride thin films using standard EBL and plasma techniques makess such compact spatial differentiators attractive for optical beam shaping, optical information processing and optomechanical sensing applications.

%%%%%%%%%%%%%%%%%%%%%%%%%%%%%%%%%%%

\section{Appendix: "TE grating"}

\begin{figure}[h]
\centering
\includegraphics[width=0.75\textwidth]{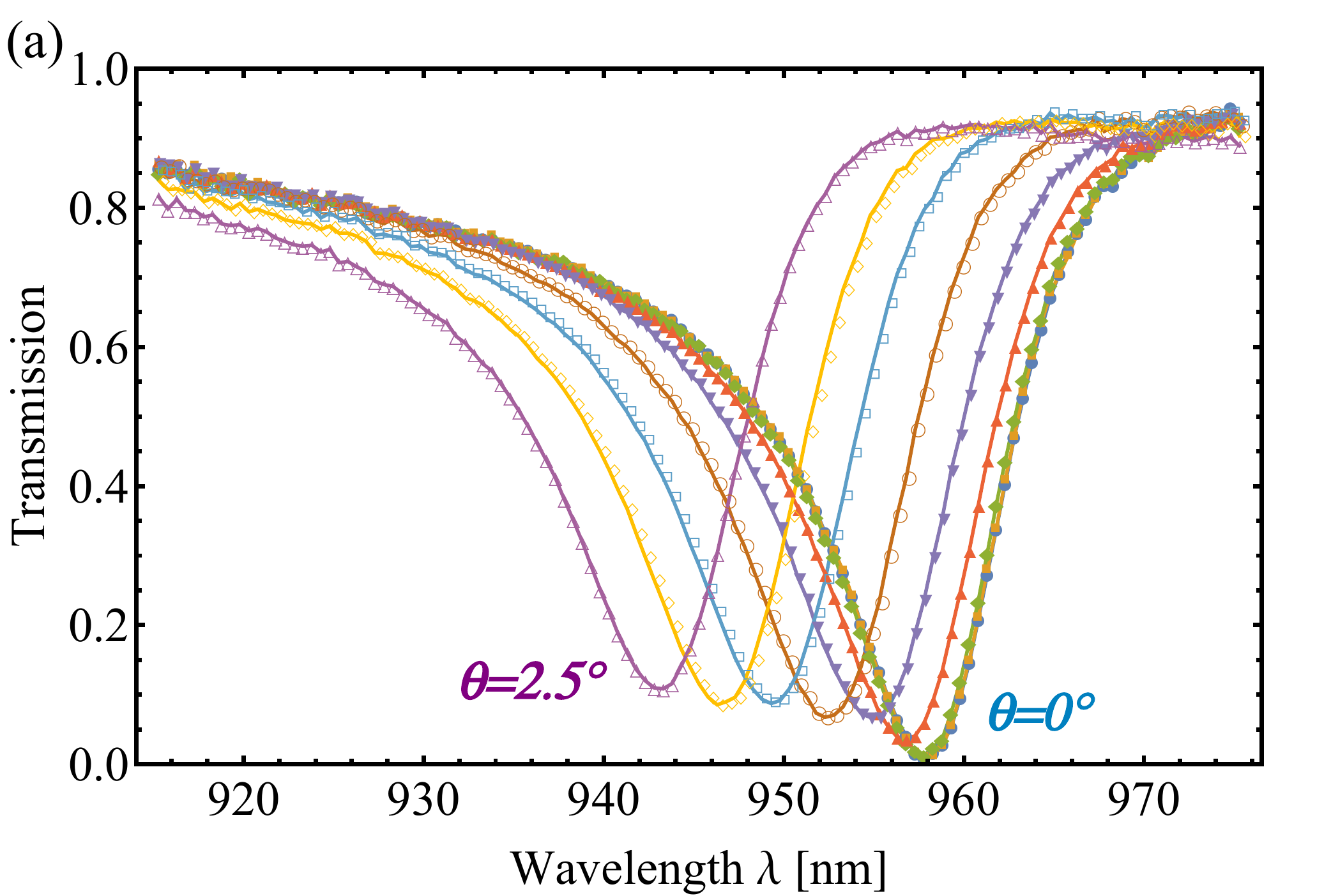}\\
\includegraphics[width=0.75\textwidth]{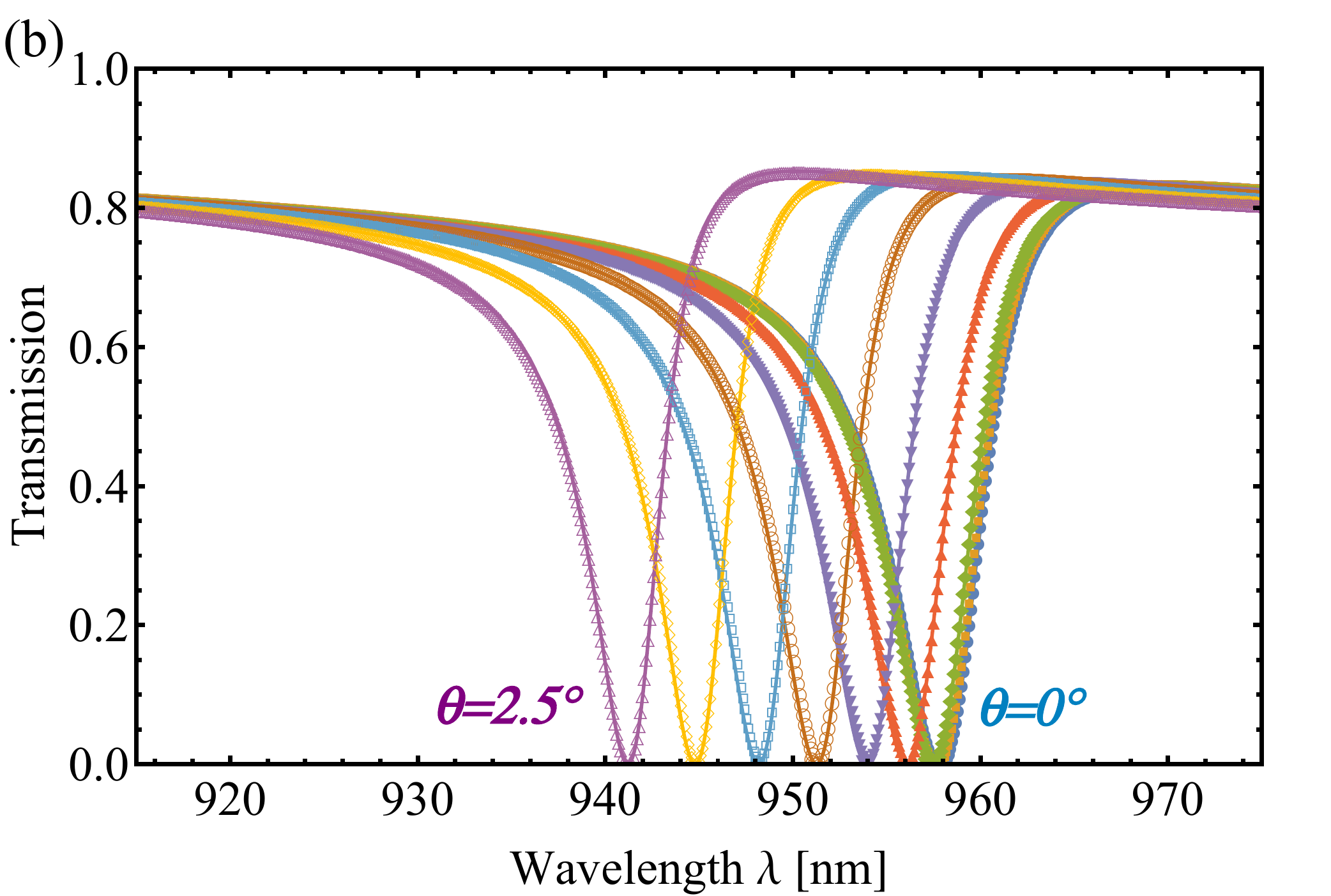}\\
\includegraphics[width=\textwidth]{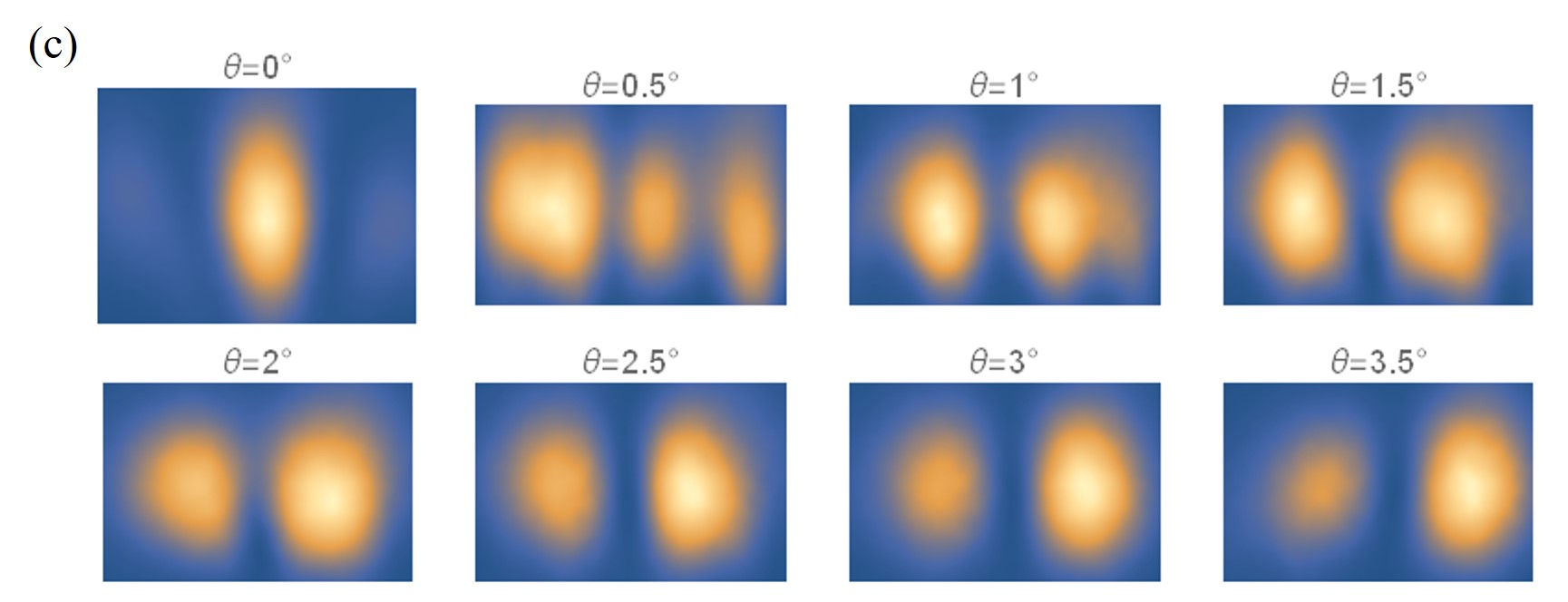}
\caption{Same as Fig.~\ref{fig:TMspectra}, but for the "TE grating" and incidence angles $\theta=0^\circ,0.25^\circ,0.5^\circ,1^\circ,1.5^\circ,2^\circ,2.5^\circ,3^\circ,3.5^\circ$.}
\label{fig:TEspectra}
\end{figure}

\begin{figure}[h]
\centering
\includegraphics[width=0.75\textwidth]{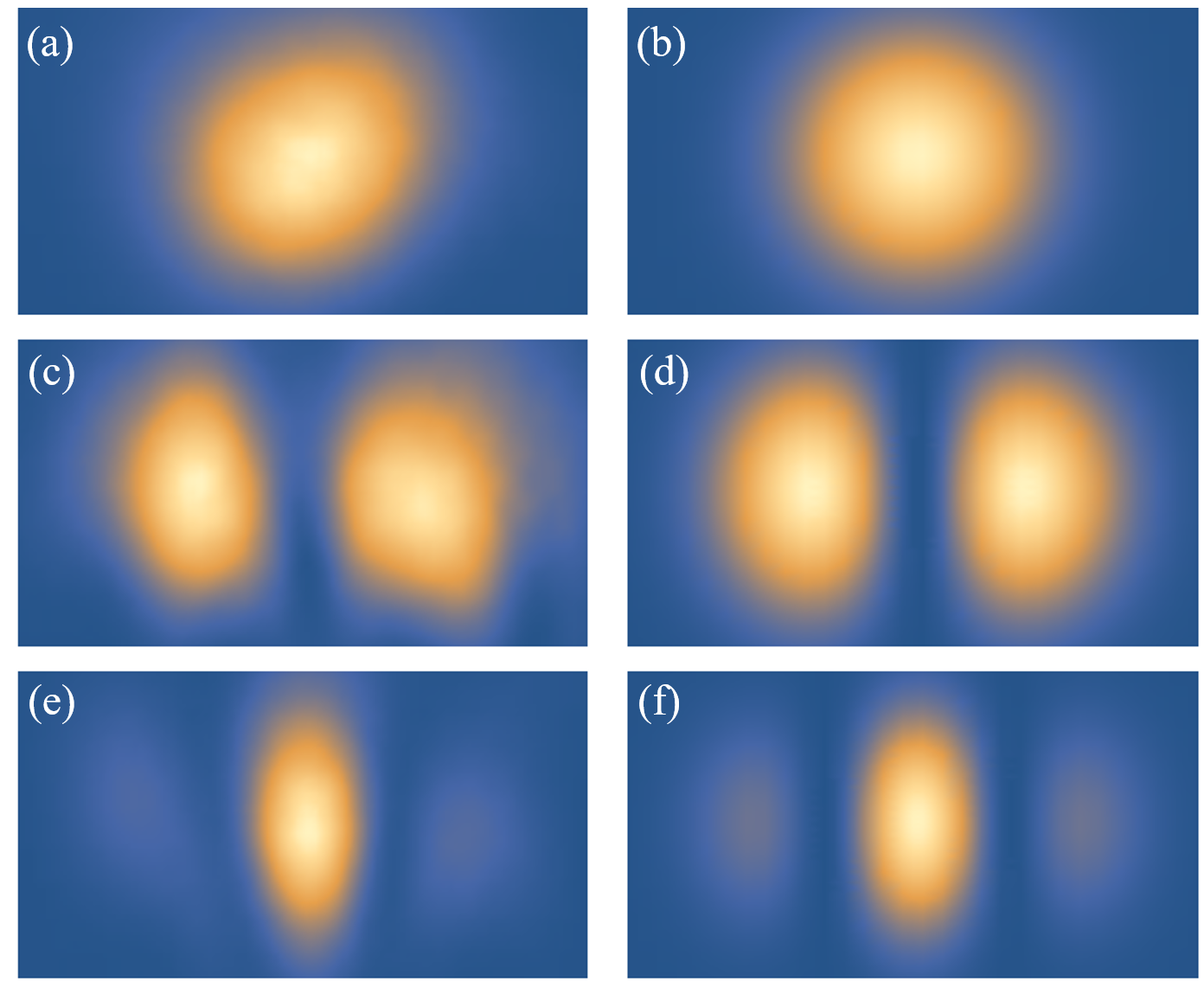}\\
\includegraphics[width=0.75\textwidth]{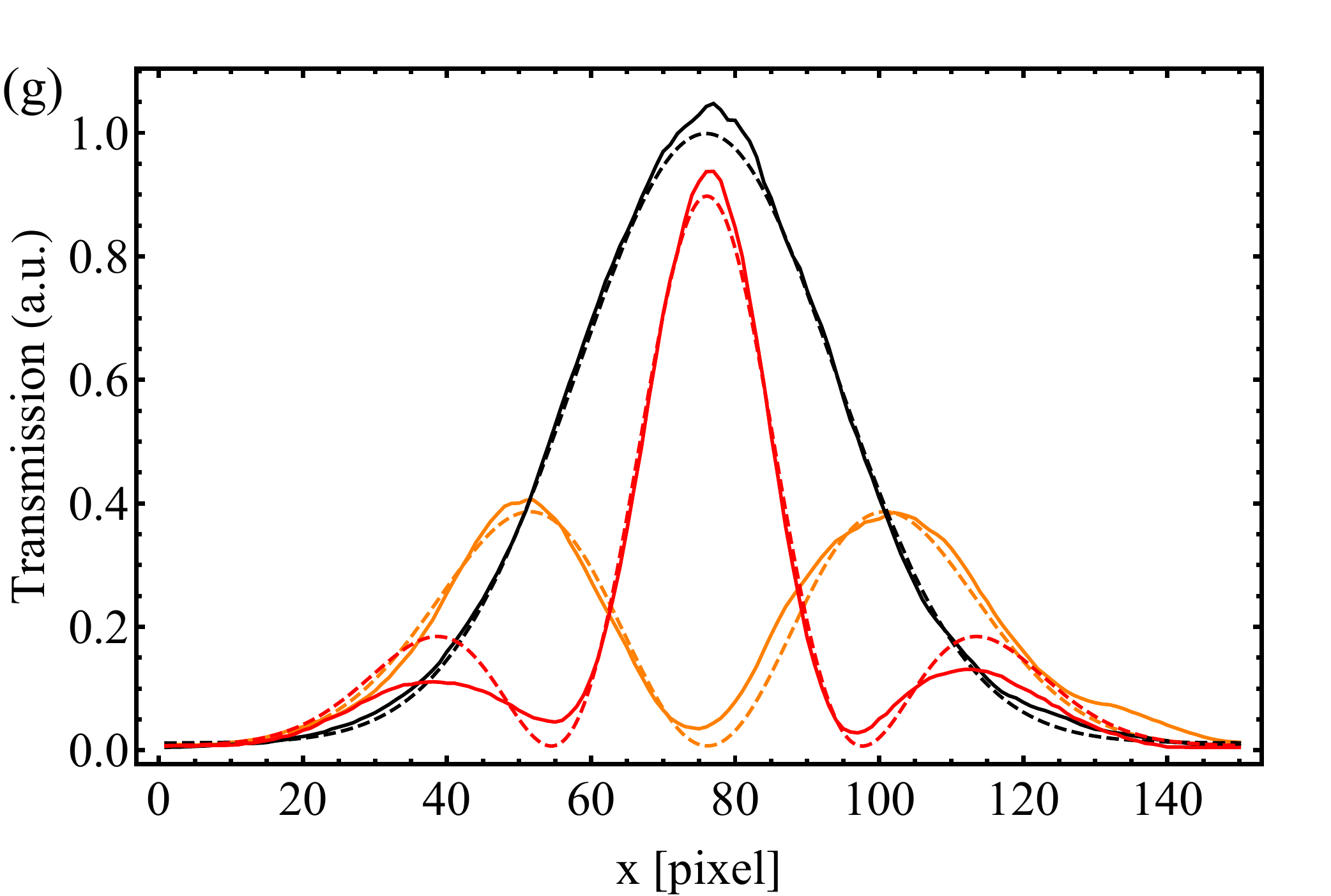}
\caption{Same as Fig.~\ref{fig:TMmodes}, but for the "TE grating". The angles of incidence in (c) and (e) are $\theta=1.5^\circ$ and $\theta=0^\circ$, respectively. The $\chi^2$ fit coefficients for the fit results shown in (b), (d) and (f) are respectively 0.989, 0.975 and 0.979.}
\label{fig:TEmodes}
\end{figure}

Experiments similar to those described in Sec.~\ref{sec:diff} were performed with the "TE grating" previously described and similar results, which we show here for completeness, were obtained. The transmission spectra measured for different incidence angles are shown in Fig.~\ref{fig:TEspectra}; the first Fano resonance occurs at 958.3 nm at normal incidence, with a minimum transmission of 1.1\%. At oblique incidence, the first Fano resonance is as expected shifted towards lower wavelengths, since the second Fano resonance occurs at a higher wavelength (outside the scanning range of the laser). First- and second-order spatial differentiation are also observed with similar behavior and quality as for the "TM grating" (see Figs.~\ref{fig:TEspectra} and \ref{fig:TEmodes}).

\section*{Funding}
Independent Research Fund Denmark.

%%%%%%%%%%%%%%%%%%%%%%% References %%%%%%%%%%%%%%%%%%%%%%%%%

\end{document}